# Dust May Be More Rare Than Expected in Metal Poor Galaxies

David B. Fisher[1,2], Alberto D. Bolatto[1], Rodrigo Herrera-Camus[1], Bruce T. Draine[3], Jessica Donaldson[1], Fabian Walter[4], Karin M. Sandstrom[4], Adam K. Leroy[5], John Cannon[6], and Karl Gordon[7]

**Affiliations**

[1]Department of Astronomy, Laboratory for Millimeter-wave Astronomy and Joint Space Institute, University of Maryland, College Park, MD 20742, USA

[2]Centre for Astrophysics and Supercomputing, Swinburne University, PO Box 218, Hawthorn, Victoria 3122 Australia

[3]Department of Astrophysical Sciences, Princeton University, Princeton, NJ 08544, USA

[4]Max-Plank Institut für Astronomie, Königstuhl 17 D-69117, Heidelberg, Germany

[5]National Radio Astronomy Observatory, Charlottesville, VA 22903, USA

[6]Department of Physics and Astronomy, Macalester College, 1600 Grand Avenue, Saint Paul, MN 55105, USA

[7]Space Telescope Science Institute, MD 21218, USA

**`Normal' galaxies observed at z>6, when the Universe was <1 billion years old, thus far show no evidence [1,2,3] of the cold dust that accompanies star formation the local Universe, where the dust-to-gas mass ratio is ~1%. A prototypical example is 'Himiko' (*z=6.6*), which a mere 840 Myr after the Big Bang is forming stars at a rate of 30-100 $M_\odot$ yr$^{-1}$, yielding a mass assembly time $M^*/SFR$~150×10$^6$ yr. Himiko is estimated to have a low fraction (2-3% of the Solar value) of elements heavier than helium (metallicity), and although its gas mass cannot be asserted at this time its dust-to-stellar mass ratio is constrained[3] to be <0.05%. The local galaxy I Zw 18, with a metallicity ~4% solar[4] and forming stars less rapidly than Himiko but still vigorously for its mass ($M^*/SFR$~1.6×10$^9$ yr)[5], is also very dust deficient and perhaps one of the best analogues of primitive galaxies accessible to detailed study. Here we report observations of dust emission from I Zw 18 from which we determine its dust mass to be 450-1800 $M_\odot$, yielding a dust-to-stellar mass ratio ≈10$^{-6}$-10$^{-5}$ and a dust-to-gas mass ratio 3.2-13×10$^{-6}$. If I Zw 18 is a reasonable analog of Himiko, then Himiko's dust mass is ≈50,000 $M_\odot$, a factor of 100 below the current upper limit. These numbers are considerably uncertain, but if most high-z galaxies are more like Himiko than like the quasar host SDSS**

J114816.64+525150.3[6], then the prospects for detecting the gas and dust in them are much poorer than hitherto anticipated.

The recent study[7] of HFLS3, a "maximum starburst" at $z=6.3$, provides an example of a galaxy with a large amount of dust ($M_{dust} \approx 10^9$ $M_\odot$), and a dust-to-gas mass ratio ($D/G \approx 0.01$) and dust-to-star mass ratio ($D/M^* \approx 0.04$) more like nearby star bursting galaxies. This galaxy has an astonishing star formation rate of $\approx 3,000$ $M_\odot$ yr$^{-1}$, and converts its gas into stars at rates 2,000 times that of typical galaxies, properties that are rare even for the gas rich high-$z$ galaxies. Frequently observations of dust and molecules in high-redshift galaxies tend to target those with bright active galaxies[8], like HFLS3 or J114816.64+525150.3. Massive galaxies like these are well known to be rare at all redshifts. For those "normal" galaxies where deep sub-mm observations have been performed, currently only upper limits for both [CII] and sub-millimeter continuum exits[1,2]. These observational limits suggest that in the first 800 Myr of galaxy evolution, galaxies with very little dust and low metallicity, like Himiko, are more typical. Results from stellar populations analyses indeed show that high redshift galaxies have very little evidence of dust extinction[9]. Understanding of the physical conditions under which stars form in these primitive systems, however, can only come from the study of local analogues.

Located at a distance of 18 Mpc[10], I Zw 18 is the archetypal star-forming very low metallicity[4] galaxy ($12+\log(O/H)=7.17$, or 1/30 Solar metallicity). I Zw 18 is gas rich[11,12,13] ($M_{HI}=2.3\times10^8$ $M_\odot$ and $M_{H2} \leq 5\times10^7$ $M_\odot$) and actively star forming[5] ($SFR=0.05\pm0.02$ $M_\odot$ yr$^{-1}$). Compared to its stellar mass, $M_* \approx 9\times10^7$ $M_\odot$, this galaxy has a very high gas fraction ($M_{gas}/M_{star+gas} \approx 2/3$). I Zw 18 is currently undergoing a starburst phase[14], and despite its active star formation there is no detected CO emission indicative of molecular gas[12] in I Zw 18. The lowest metallicity detection of CO was recently reported in the dwarf galaxy WLM[15] ($12+\log(O/H) \approx 7.8$ or 1/8 Solar). Unlike WLM, I Zw 18 has a very active star-forming environment which likely photodissociates CO. These properties mean that it is among the closest analogues to primitive high-redshift galaxies, although I Zw 18 contains a more significant population of evolved stars than may be found in the early Universe.

Note that the difference in stellar mass between I Zw 18 (or any local low metallicity galaxy) and observed high-redshift galaxies is large. Because of the smaller potential well, identical starbursts can in principle more easily drive dust and metal-enriched gas out I Zw 18 than out of Himiko for example. Nonetheless, I Zw 18 and galaxies like it remain our best candidates for the study of metal-poor, starbursting environments.

Using Herschel PACS, we measure the flux of I Zw 18 to be 21.1 mJy at 100 $\mu$m, with an uncertainty in flux (calculated by placing apertures randomly in the map) of ±2.9 mJy (signal-to-noise, $S/N \sim 7$) and a calibration uncertainty of 10% (±2.1 mJy). At 160 $\mu$m we measure a flux of 5.6 mJy, and a flux uncertainty in the map of ±1.3 mJy ($S/N \sim 4$) and calibration uncertainty of ±0.6 mJy. (Maps are shown in Fig. 1 and our procedure is discussed in the Supplementary Information.) Together with these detections, we use a number of ancillary data sources to construct a full infrared SED for modeling the dust and star formation. We employ data from Spitzer covering 3.6, 4.5, 5.8, 8.0, 24 and 70 $\mu$m[16,12] as well as an IRS spectrum[14] (see Supplementary Information).

The dust mass we determine from models (shown in Figure 2) with mixed dust grain temperatures is $M_{dust}$ = $912^{+912}_{-456}$ $M_\odot$ (see Supplementary Information for a complete discussion of the uncertainties). Modified black body models are also commonly used for fitting dust SEDs, although they yield unrealistically low dust masses because of the assumption of a single temperature. For comparison, a modified blackbody model with $F_\nu \propto \nu^{1.5} B_\nu(T_{dust})$ and mass-emissivity $\kappa_{200}$ = 6.37 cm$^2$ g$^{-1}$, reproduces the flux at 70 and 160 $\mu$m for $T_{dust}$ = 70 K and $M_{dust} \cong$ 250 $M_\odot$. We find that a significant mass of cold dust with $T$>15 K would be detected by our 160 $\mu$m flux (see Supplementary Information). Independent of the assumed model, the dust mass necessary to explain the SED observed in I Zw 18 is extremely low.

Under the simplest assumptions decreasing the amount of heavy elements, which constitute the dust particles, results in a proportional decrease in the dust-to-gas ratio. This scenario is frequently assumed in cosmological models of star formation[17]. Observational constraints for this relationship in the early Universe are scarce. I Zw 18 provides a probe of such environments, and our measurements directly constrain the relationship between dust-to-gas ratio and metallicity at very low metallicities[12,18,19] (12+log($O/H$)≤ 8). We find that I Zw 18 falls roughly two orders of magnitude below the linear correlation between metallicity and dust-to-gas ratio (Figure 3). The distance to the linear relation is very significant, approximately four times larger than the spread in the data, and much larger than the errorbars on the measurement. Using dust-to-gas ratios measured only in the IR emitting region typically results in a linear relationship even at low metallicity[18], 12+log($O/H$)~8, but not in the case of I Zw 18 where $DGR_{local}$ is still a factor of 38 below the linear relationship.

I Zw 18 stands out in the local galaxy population because it has an environment that is both star bursting and also lacking heavy elements. Our results suggest that in starbursting galaxies with very low metallicities the dust-to-gas ratio is determined by more than just the availability of heavy elements. The essentially dust-free character of nascent galaxies[1,2,3], like Himiko, therefore likely reflects a combination of low metallicity and the balance of the dust production and destruction mechanism in a starburst environment, which act together to keep the dust-to-gas ratio very low in these galaxies.

These are properties that are common at very high redshift ($z$>6), and consequently we expect those primitive galaxies to exhibit very low dust masses compared to their star formation rates and stellar masses. The ratio of $M_{dust}$-to-$SFR$ (Figure 4) in I

Zw 18 is more than two orders of magnitude lower than typical in local galaxies, and an order of magnitude lower than observed in the $z$=2-3 starbursts. Even when normalized by its stellar mass or *SFR*, the dust mass of I Zw 18 is extremely small compared to both local and $z$=2-3 galaxies. In Figure 4 we show that the difference between interpreting the dust mass of Himiko using the upper limit on the 1.2 mm flux[3], or using I Zw 18 as an analogue has very significant impact on the inferred properties. With considerable uncertainty, we can scale the stellar mass of Himiko with the dust-to-stellar mass ratio of I Zw 18, and place it on Fig. 4 with a dust mass of ≈50,000 $M_\odot$. If the dust temperature is 40 K[3], we calculate Himiko would have a flux density of 0.5 µJy at Earth at 260 GHz, which would require several tens of days of integration with the complete Atacama Large Millimeter Array (ALMA) to detect it. Maximum starbursts like HFLS 3 are very rare, even in the early Universe[7] (≈1 Gpc$^{-3}$), whereas blue dust-poor "drop-out" galaxies are much more common at high redshifts[9] ($10^{-3}$ Mpc$^{-3}$). This implies that the ISM of I Zw 18 may indeed be representative of the primitive galaxy population in the early Universe. If this is the case the prospects for detecting dust emission at $z$>6 will likely be limited to the unusual evolved sources like HFLS3 and J1148+5251.

**Supplementary Information** is linked to the online version of the paper at www.nature.com/nature.

**Acknowledgements** DBF, ADB, and RHC acknowledge support from University of Maryland and the Laboratory for Millimeter Astronomy and NSF-AST0838178. ADB acknowledges partial support from a CAREER grant NSF-AST0955836, NSF-AST1139998, and from a Research Corporation for Science Advancement Cottrell Scholar award. BTD acknowledges partial support from NSF-AST1008570. J.M.C. is supported by NSF grant AST-1211683. KMS acknowledges support from a Marie Curie International Incoming fellowship. PACS has been developed by a consortium of institutes led by MPE (Germany) and including UVIE (Austria); KU Leuven, CSL, IMEC (Belgium); CEA, LAM (France); MPIA (Germany); INAF-IFSI/OAA/OAP/OAT, LENS, SISSA (Italy); IAC (Spain). This development has been supported by the funding agencies BMVIT (Austria), ESA-PRODEX (Belgium), CEA/CNES (France), DLR (Germany), ASI/INAF (Italy), and CICYT/MCYT (Spain). We are thankful to both the referees and editors for helpful comments on this manuscript.

**Author Contributions:** DBF and ADB wrote the text of both the proposal and this manuscript. DBF and RHC performed detailed calculations. JD reduced the Herschel data. BTD modeled the SED. AKL and FW obtained and reduced CO observations. JC obtained the Hα flux for I Zw 18. All authors participated in discussion of results and helped with revision of the manuscript.

**Author Information** Correspondence and requests for materials should be addressed to DBF (dfisher@swin.edu.au) or AB (bolatto@astro.umd.edu).

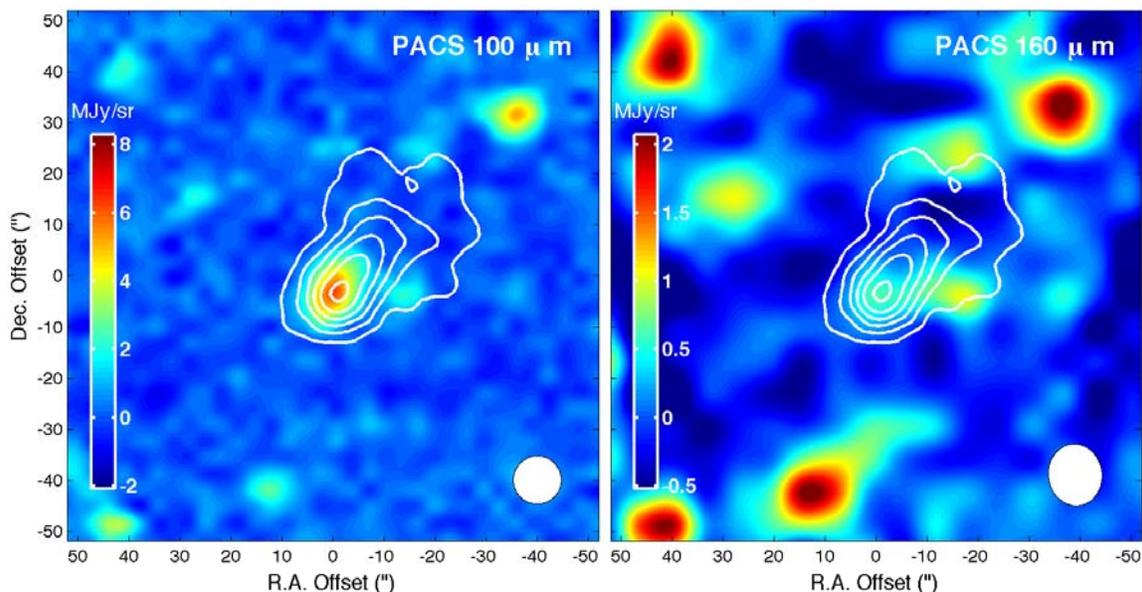

FIG. 1.— **The 100 $\mu m$ and 160 $\mu m$ images of I Zw 18.** In this figure we show the first λ = 100 $\mu$m far-infrared (FIR) detection of dust emission in I Zw 18, and the marginal detection of I Zw 18 at 160 $\mu$m. These new observations (PID: OT_dbfisher_1) were obtained with *Herschel* PACS. White contours show N(HI) = 0.7, 1.4 and 5 × $10^{20}$ cm$^{-2}$ from the Very Large Array (VLA) map[11] . The beam size of the HI map is 8.8×8.3 arcsec. For display purposes, the pixel size of the infrared maps is resampled to match the pixel size of the HI map. At 100 $\mu$m I Zw 18 is clearly detected, and well matched to the center of the HI gas contours. The emission we detect in both far infrared filters is contained within a small region (15" or 1.3 kpc). We note that the off target peaks in the 160 $\mu$m map are not noise, they are all coincident with peaks in the 100 $\mu$m map, and are therefore most likely background targets. At 160 $\mu$m we detect emission at the 3σ level that is consistent with the peak of both HI and the IR emission at 24, 70 and 100 µm, and which we attribute to I Zw

18. The peak and extent of the emission in our images is coincident with that of the Hα emission from HST images[5], and also the peak of HI emission[11].

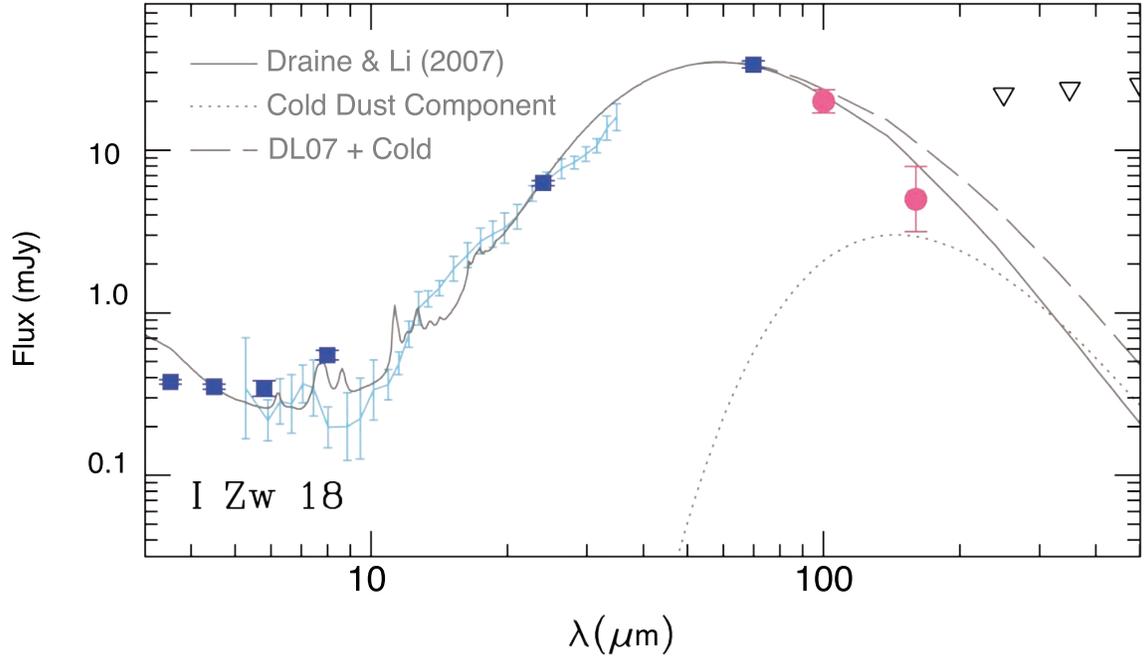

FIG. 2.— **The far infrared spectral energy distribution of I Zw 18.** Blue squares represent fluxes measured by IRAC and MIPS onboard *Spitzer*[16,12]. The blue line is a smoothed IRS spectrum[14]. The open triangles are upper limits from *Herschel* SPIRE. The red circles show our new *Herschel* PACS data at 100 and 160 μm. Error bars represent 1σ uncertainties. We fit the SED (spectral energy distribution) with models based on a mixture of carbonaceous and silicate dust grains[18], which assume a distribution of grain sizes matching that of the Milky Way. This is a commonly adopted dust model, allowing direct comparison to dust masses in the literature, and is not expected to introduce large errors in the dust mass estimate[18]. Most of the modeled dust is heated by a single starlight intensity ($U_{min}$), but a fraction is heated by a power-law distribution of intensities, with an adjustable slope (α) and upper cut-off ($U_{max}$). The best-fit dust model to the SED of I Zw 18, shown in Figure 2, returns the following values: α = 2.4, $U_{min}$ = 100, and <U> ≈ 200, for $U_{max}$ = $10^7$ and no polycyclic aromatic hydrocarbons, with $U$ in units of the radiation field in the vicinity of the Sun. The solid grey line represents the best-fit dust model. The dotted line represents a cold (T=20 K) dust model with a mass of 1000 $M_\odot$ and κ that is proportionate to $v^{1.5}$. The dashed line represents the linear combination of the cold component and the dust model. We find that including the cold component of dust increases the discrepancy between the model and the 160 μm flux to 2.3σ. Our assumption of a factor of two uncertainty on the dust mass, therefore, accounts for the possibility of the cold component.

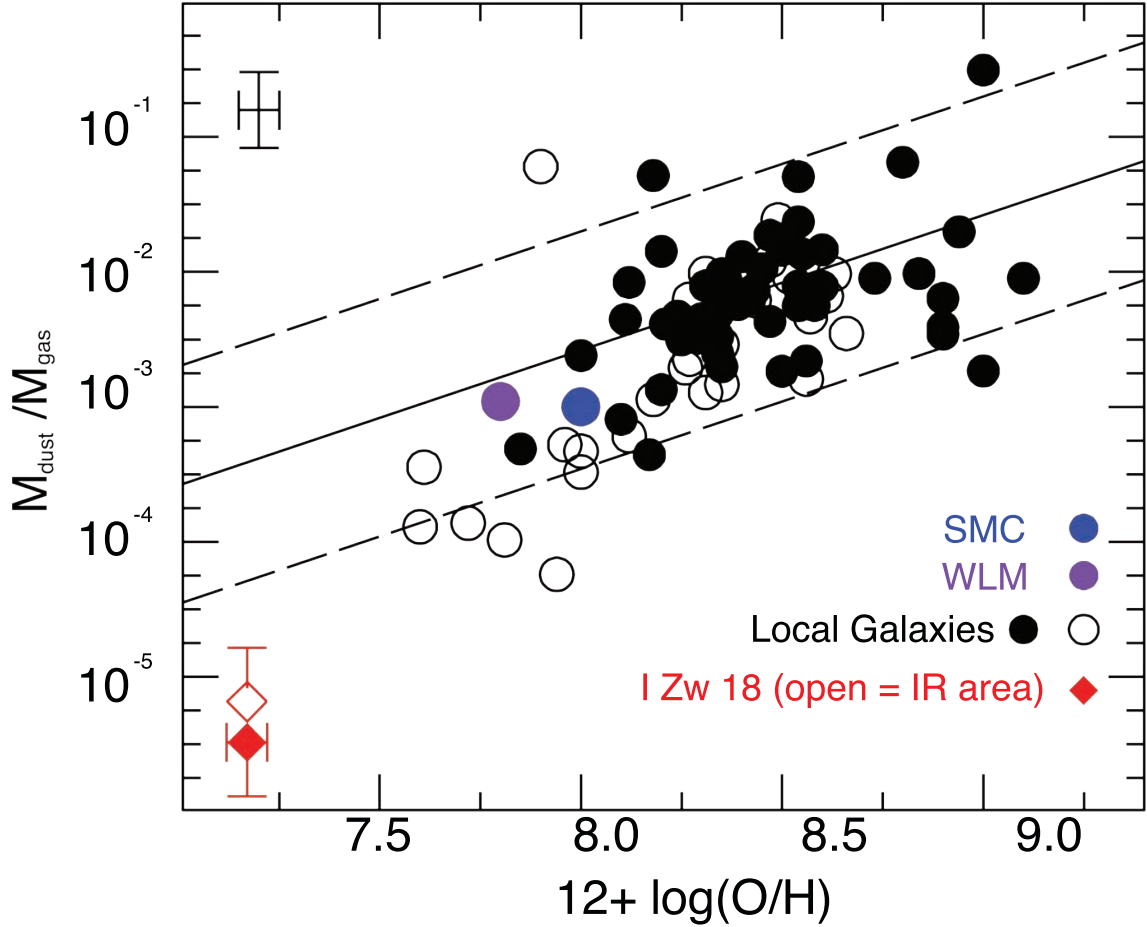

FIG. 3.— **The dust-to-gas ratio of galaxies compared to metallicity for local galaxies and I Zw 18.** The Local Galaxy sample consists of dust-to-gas ratios from two recent papers[21,19] the sample is representative of typical disk and dwarf galaxies in the local Universe. For some galaxies in the local sample either HI or $H_2$ are not available in the literature. In this case we estimate the total gas mass with an empirical correlation relating molecular to atomic gas, *M($H_2$) = 0.008 M(HI)$^{1.2}$*. Those galaxies are plotted as open circles. We also include two nearby, well-known low metallicity galaxies WLM[15] (purple circle) and the SMC[20] (blue circle). The error bars in the upper left corner represent the median *2σ* for the Local galaxy sample. A common assumption is that dust-to-gas ratio in galaxies scales linearly with metallicity[17]. We therefore show a linear bisecting line set to match the local galaxy sample, the dashed lines represent the ±2σ root-mean-square scatter around this bisector. I Zw 18 is a clear outlier from this correlation. Note the distinction between both SMC and WLM from I Zw 18 is the starbursting nature of I Zw 18. If I Zw 18 is representative of starbursting low-metallicity environments, this implies that the dust mass is lower than we may expect in primitive galaxies of the early Universe.

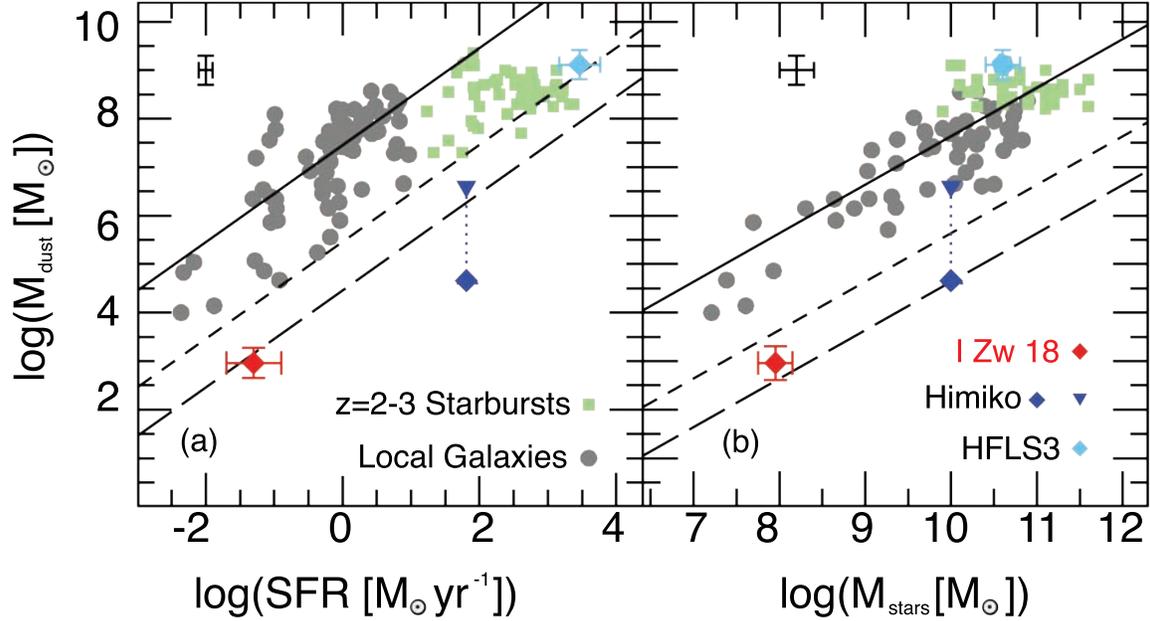

FIG. 4.— **Dust mass versus star formation rate and stellar mass for local disks, high-redshift star bursts and I Zw 18.** Here we compare dust masses to star formation rates (a) and stellar masses (b) of a sample of normal and starbursting galaxies[21,19,22,23,24], including I Zw 18. The definitions of symbols are the same in both panels. The red diamond represents I Zw 18. Squares represent $1<z<3$ starbursts[25]. The blue triangle represents the current upper-limit on the dust mass of Himiko, the blue diamond shows the dust mass of Himiko if it has a similar dust-to-stellar mass ratio as I Zw 18. Solid lines represents linear relationships matching the local galaxies, the dashed lines represent 1/100 (short dashes) and 1/1000 (long dashes) of the local sample. The error bars in both panels represent the median $2\sigma$ error bar for the sample. Note that very significant differences exist between I Zw 18 and the other starbursting systems shown here; the distant starbursts have stars formation rates five orders of magnitude higher than I Zw 18, and many of them have solar metallicity[25]. I Zw 18 is seen to be a clear, extreme, outlier toward lower ratios of dust-to-SFR and dust-to-gas when compared to typical nearby galaxies, and the dust mass is even lower per unit SFR than star bursting galaxies. As indicated by the extreme difference in the current upper limit of Himiko, and the predicted dust mass using the dust-to-star mass ratio of I Zw 18, those observations of the highest redshift galaxies may be significantly over estimating the dust mass.

**Supplementary Information**

*Flux Measurements:* The data are reduced in HIPE 8.2[26] using standard methods. To measure the flux we choose a circular region centered on the peak of the 100 μm image. The radius is chosen to correspond to the point at which the 100 μm surface brightness is equivalent to the background noise (15"). The beam size of our maps is 8" and 12" at 100 μm and 160 μm respectively. The background is removed by fitting a 2 dimensional plane to the map. We apply an aperture correction, according to the Herschel documentation, of 1.238 and 1.383 to the 100 and 160 μm photometry respectively.

The measurement uncertainty on the *Herschel* fluxes is a combination of an assumed zero point uncertainty and also the uncertainty introduced from the background noise. Based on our own comparison of fluxes from PACS and MIPS, we assume an estimate of 10% for the calibration uncertainty. To measure the uncertainty in the maps we generate a set of apertures equal in size to our flux measurement (15") that are randomly placed to fill the map, excluding the region we identify as I Zw 18. We find that, on average, the expected flux of a 15" aperture is 2.9 mJy in the 100 μm map and 1.26 mJy in the 160 μm map. We also note that there is a source of emission that is very near I Zw 18 in both the 100 and 160 μm maps. We cannot know for certain if this is part of the I Zw 18 or not, so we compute the flux in a flux of 24" in radius, which includes the extra emission. We find that this introduces an uncertainty of 2.7 mJy at 100 μm and 1.6 mJy at 160 μm. This is comparable to the uncertainty from randomly placed apertures, and therefore is accounted for in our error bars. In our maps the calibration and measurement uncertainties contribute comparable amounts to the total uncertainty. The total uncertainty is therefore computed by adding in quadrature the calibration and image uncertainties.

We use Hα fluxes obtained from HST mapping[5]. The Hα flux is measured in a continuum subtracted image, and integrated over a region that is collocated with the IR emitting region. For more details about the Hα we direct the reader to its original publication[5]. We convert this flux to a star formation rate with a recent calibration[27]. We use VLA mapping[11] to measure the HI mass of the galaxy. The CO(1-0) measurements of I Zw 18 return an upperlimit[12] of the molecular gas mass of $M_{H2}$ ≤ 4.5×10$^5$ $M_\odot$, using a galactic conversion factor for CO flux-to-molecular gas mass. Alternatively, we can estimate the molecular gas mass by assuming that I Zw 18 follows a similar star formation law as other galaxies[28], such that $M_{mol}$ ≈ (2×10$^9$) × $SFR$, and thus $M_{mol}$≈10$^8$ $M_\odot$. Indeed when we adopt the metallicity dependent conversion factor[29] the upper limit to the molecular gas mass is $M_{H2}$ ≤ 5×10$^7$ $M_\odot$. In this paper we use 5×10$^7$ $M_\odot$ as the molecular gas mass of I Zw 18. We note that this amounts to only 30% of the total gas mass, and has, therefore, a very small effect on our main result (in Figure 3).

We find that the flux in the 4.5 $\mu m$ is heavily dominated by the star light emission and not by Br α line emission. We determine this by scaling the Hα flux by 0.0273 (10$^4$ K Case B[30]). We find that the Br α line accounts for roughly ≈2% of the flux in IRAC band 2.

*Fitting the Dust Model to the SED:* In the fits the IRAC, MIPS and PACS data points are weighted by a factor of 5 and the IRS data points are given a weight of 1; this is done to produce and even weighting across all wavelength ranges. Otherwise the software would only seek to fit the IRS data. Also, all points are weighted by the inverse of the uncertainty squared. In Figure 2 there is a slight difference between the flux at 8 $\mu$m using either IRAC or IRS data. One simple explanation is that this discrepancy is due to imperfections in the IRAC channel 3 and 4 zero points. In light of this, we do not include these channels in the fit. Instead the fit is constrained by the IRS data alone over these wavelengths. Inspection of published IR SEDs of other galaxies shows that these discrepancies are not extremely uncommon. For completeness, we also fit the full SED, including both IRAC bands 3 & 4 and all other data. We find that this does not significantly affect the dust mass, $M_{dust}$(all data) = 881 $M_\odot$.

We set $q_{PAH}$=0 in the model. This is consistent with previous observations of low metallicity galaxies. We note that the PAH mass in the $qPAH$=0 model is not exactly zero, rather the size distribution of carbonaceous grains is set to go to zero smoothly, and therefore has a very small amount of material in the range of sizes that correspond to PAH grains. (It is for this reason that a few PAH lines are present in our dust model SED.)

The dust mass uncertainty comes from three sources: errors in the flux measurements, the scatter in the fit values, and the systematic uncertainty in the model. We estimate the uncertainty on the dust mass from the flux measurements by combining the error bars shown in Fig. 2. The uncertainty is dominated by the error in the 160 $\mu$m measurement. We ran 20 independent realizations of the model, with Gaussian noise added, and the resulting scatter in dust mass is ≈20%. Including sub-mm fluxes (λ > 160 μm) in low metallicity dwarf galaxies, and including a cold dust population in models, systematically increases the dust mass[19] by roughly a factor of ~ 1.5. In Fig. 2 open triangles represent upper limits measured from Herschel. Only upper limits exist for the sub-mm flux of I Zw 18. Conversely, assuming a Milky Way dust distribution may overestimate the dust mass[31] in a low metallicity environment by a factor of 2. For physical dust models, differences in assumptions about grain size distributions, compositions and mass emissivities introduce uncertainty in the dust mass at the factor of two level[18]. To account for these possible systematic uncertainties we assume that the dust mass is uncertain by a factor of 2.

In Figure 2 we show the effect of including a dust component with T=20 $K$ and mass $M_{cold}$=1,000 $M_\odot$. The fitted dust model is already higher than the flux at 160 $μm$ by 1.3σ, and including this cold dust component increases the discrepancy to 2.3σ. Nonetheless, a much colder dust component would be impossible to reject with the existing data, and in fact a *T*=10 K component of $M_{cold}$=10,000 $M_\odot$ would be undetectable in our observations. As stated above, such a component would likely be outside of the IR emitting region, given the high radiation fields. Recent evidence[32] from HST/Cosmic Origins Spectrograph suggests that in the HI envelope the metallicity is likely even lower than that in the center of the galaxy, with "pockets of pristine gas, free of any metals." Such observations suggest that it is unlikely that significant amounts of cold dust exist in the HI envelope surrounding the IR emitting region.

The lower cutoff for the radiation field in I Zw 18 is $U_{min}$=100. This high value, five times larger than the largest $U_{min}$ typically found in nearby galaxies[18], suggests an absence of quiescent regions in I Zw 18 over a region that is roughly 3 kpc in diameter. Typical values of *<U>* are 0.7-30 in samples of nearby galaxies[18]. The average dust grain in I Zw 18 experiences a radiation field that is many times stronger than in a typical galaxy, and even more extreme than in distant starbursts[25].

*IRS Spectrum:* The coverage of the IRS spectrum is a slit that runs through the center of IR region[13]. The slit is only ~4" wide and therefore does not observe all the extended, low surface brightness emission. Nonetheless, the slit easily includes the brightest parts of the H*α* map.

All metallicities are displayed in the Pilyugin & Thuan (2005) calibration.[33]

26. Ott, S. Astronomy Society of the Pacific Conference Series Vol 434 Astronomical and Data Analysis Software and Systems XIX ed. Y. Mizumoto, K.I. Morita, & M. Ohishi 139-142 (2010)